\documentclass[aps,pre,twocolumn,showpacs,floatfix,superscriptaddress]{revtex4}

\usepackage{graphicx}
\usepackage{epsfig}
\usepackage{epstopdf}
\usepackage{bm}

\begin{document}

% Title Page
\title{Nonparametric Maximum Entropy Estimation on Information
  Diagrams}
%\title{Nonparametric Maximum Entropy Estimation Given Information-Theoretic Constraints}

\author{Elliot A. Martin}
\affiliation{Complexity Science Group, Department of Physics and
  Astronomy, University of Calgary, Calgary, Alberta, Canada, T2N 1N4}
\author{Jaroslav  Hlinka}
\affiliation{Institute of Computer Science, The Czech Academy of Sciences, Pod vodarenskou vezi 2, 18207 Prague, Czech
  Republic}
\affiliation{National Institute of Mental Health, Topolov\'{a} 748, 250 67 Klecany, Czech Republic}
\author{Alexander Meinke}
\affiliation{Complexity Science Group, Department of Physics and
  Astronomy, University of Calgary, Calgary, Alberta, Canada, T2N 1N4}
\author{Filip D\v{e}cht\v{e}renko}
\affiliation{Institute of Psychology, The Czech Academy of Sciences, Prague, Czech Republic}
\affiliation{Institute of Computer Science, The Czech Academy of Sciences, Pod vodarenskou vezi 2, 18207 Prague, Czech
  Republic}
\author{J{\"o}rn Davidsen}
\affiliation{Complexity Science Group, Department of Physics and
  Astronomy, University of Calgary, Calgary, Alberta, Canada, T2N 1N4}

\date{\today}

\begin{abstract}
  Maximum entropy estimation is of broad interest for inferring
  properties of systems across many different disciplines. In this
  work, we significantly extend a technique we previously introduced
  for estimating the maximum entropy of a set of random discrete
  variables when conditioning on bivariate mutual informations and
  univariate entropies. Specifically, we show how to apply the concept
  to continuous random variables and vastly expand the types of
  information-theoretic quantities one can condition on. This allows
  us to establish a number of significant advantages of our approach
  over existing ones. Not only does our method perform favorably in
  the undersampled regime, where existing methods fail, but it also
  can be dramatically less computationally expensive as the
  cardinality of the variables increases. In addition, we propose a
  nonparametric formulation of connected informations and give an
  illustrative example showing how this agrees with the existing
  parametric formulation in cases of interest. We further demonstrate
  the applicability and advantages of our method to real world systems
  for the case of resting-state human brain networks. Finally, we show
  how our method can be used to estimate the structural network
  connectivity between interacting units from observed activity and
  establish the advantages over other approaches for the case of phase
  oscillator networks as a generic example.
\end{abstract}

\pacs{89.75.Hc, 89.70.Cf, 05.45.Tp, 87.18.Sn}

\maketitle

\section{Introduction}

% \begin{itemize}
% \item Maximum entropy - particularly for statistical inference -
%   Jaynes
% \item Decoding Neurons, genetic networks, other examples --- network
%   inference/connected informations
% \item typical methods using max ent --- conditioning on
%   cross-correlations or distributions
% \item motivate why we want to use information theoretic quantities
% \item Yeung has extended information theory to set theory *** Check
%   original source for diagram construction
% \item How information diagrams are constructed
% \item hard to maximize entropy conditioned on mutual information and
%   entropies using Lagrange multipliers --- transcendental equations
% \item define terms, entropies, conditional entropies, mutual
%   information, conditional mutual information, multivariate mutual
%   information
% \item Summary of paper, what each section entails
% \end{itemize}

% {\bf I1: The tendancy of systems to approach a state of maximum
%   entropy has been known for a long time, and is enshrined in the
%   second law of thermodynamics. Statistical physics tells us that the
%   most probable state of a system is the one that maximizes it's
%   entropy. Jaynes later showed that one can consider statistical
%   mechnanics as a general form of statistical inference, where the
%   least biased estimate is the one which maximizes the entropy.}

Statistical mechanics is based on the assumption that the most
probable state of a system is the one with maximal entropy. This was
later shown by Jaynes~\cite{Jaynes1957::PRa} to be a general property
of statistical inference --- the least biased estimate must have the
maximum entropy possible given the constraints, otherwise you are
implicitly or explicitly assuming extra constraints. This has resulted
in maximum entropy methods being applied widely outside of traditional
statistical physics.

Uses of maximum entropy methods can now be found in such diverse
settings as neuroscience~\cite{Schneidman2006::Nat},
genetics~\cite{Lezon2006::PNAS}, and inferring multidrug
interactions~\cite{Wood2012::PNAS}. These methods typically condition
on quantities such as cross-correlations, which are not capable of
detecting nonlinear relationships. Alternatively, one could condition
on the probability distributions of subsets of
variables~\cite{Schneidman2003::PRL, Stephens2010::PRE}, but these can
be hard to estimate accurately. In either case, the computational
costs quickly become prohibitive as the number of discrete states the
random variables can take on increases (i.e. the cardinality of the
variables increases).

In order to overcome these difficulties we propose conditioning on
information-theoretic quantities, such as entropies and mutual
informations.  For example, the bivariate mutual information can
detect arbitrary interactions between two variables, and is only zero
when the variables are pairwise independent~\cite{Cover2006}.  At the
same time these measures can often be accurately estimated at samples
sizes too small to accurately estimate their underlying probability
distributions~\cite{Nemenman2011::Ent}.

In theory, conditioning on information-theoretic quantities can be
accomplished using Lagrange multipliers. However, while this results
in relatively simple equations when conditioning on moments of
distributions, conditioning on information-theoretic quantities
results in transcendental equations --- making them much harder to
solve. The absence of techniques to efficiently calculate the maximum
entropy in these cases is conspicuous; conditioning on the univariate
entropies alone is equivalent to assuming the variables are
independent, a widely used result, but a generalisation to a wider
array of information-theoretic terms has not been forthcoming to the
best of our knowledge. In~\cite{Martin2015::PRL} we introduced a
method to address this issue using the set-theoretic formulation of
information theory, but only when conditioning on bivariate mutual
informations and univariate entropies for discrete random variables.

Here, we significantly extend this technique and provide relevant
mathematical proofs. Specifically, we show how to apply the concept to
continuous random variables and vastly expand the types of
information-theoretic quantities one can condition on. To establish
the practical relevance of our maximum entropy method, we show that it
can successfully be applied in the undersampled regime, and that the
computation time does not increase with the cardinality of the
variables --- in fact we show our method can be computed much faster
than other techniques for cardinalities greater than 2. These are two
issues that severely limit current maximum entropy methods as noted
in~\cite{Yeh2010::Ent}. Inspired by this, we construct a noparametric
estimate of connected informations introduced
in~\cite{Schneidman2003::PRL}, which are used to estimate the
relevance of higher-order interactions in sets of variables. Previous
techniques to estimate connected informations can also be hampered by
insufficient sampling, as well as become computationally intractable
for larger cardinality variables.

We are also able to use our method to help resolve an outstanding
issue of applying maximum entropy models to functional magnetic
resonance imaging (fMRI) data, where past methods showed that pairwise
measurements were a good representation of the data only when it was
discretized to two states~\cite{Watanabe2013::NatCom}. Here we show
that discretizing to larger cardinalities does not appreciably affect
results from our method, though it does for methods only conditioning
on the first two moments of the variables. This indicates that
nonlinear relationships are important for reconstructing this data.

As a final application we show how our method can be used to infer
structural network connections. Inferring networks from dynamical time
series has seen much attention~\cite{Timme2014}, with applications in
such diverse fields as neuroscience~\cite{Eguiluz2005::PRL},
genetics~\cite{Margolin2006::BMCbioinfo}, and the
climate~\cite{Runge2015::NC}, as well as for generic coupled
oscillators~\cite{Tirabassi2015::SR}. Our maximum entropy estimate
allows for the inference of the conditional mutual information between
every pair of variables conditioned on all remaining considered
variables. This has previously been used in~\cite{Frenzel2007::PRL} to
detect causal connections with some success, though it becomes
increasingly hard to estimate as the number of variables and their
cardinality go up --- due to the exponentially increasing phase
space. It has also been noted that there are fundamental issues in the
implementation of reconstructing the underlying time
graphs~\cite{Runge2015}.  Our method can help overcome the sampling
issue by not estimating the conditional mutual informations directly,
but by finding the values of the conditional mutual information
consistent with the measured pairwise mutual informations and
univariate entropies when the joint entropy is maximized.

The outline of our paper is as follows. In Sec.~\ref{Sec:Method} we show
how one can vastly increase the types of information-theoretic
quantities that one can condition on using the method we introduced
in~\cite{Martin2015::PRL}, as well as extend the method to continuous
variables. Next, in Sec.~\ref{Sec:Proofs} we prove various properties
relevant to the method. Finally, in Sec.~\ref{Sec:App} we illustrate
pertinent features of our method, and discuss various applications.

% ***Ilya's ARCANE technique essentially looks at the max ent state of
% triplets...

\section{Method}
\label{Sec:Method}

\begin{figure}[!h]
%%\begin{figure}[htbp]
  \begin{center}
    \includegraphics*[width=\columnwidth]{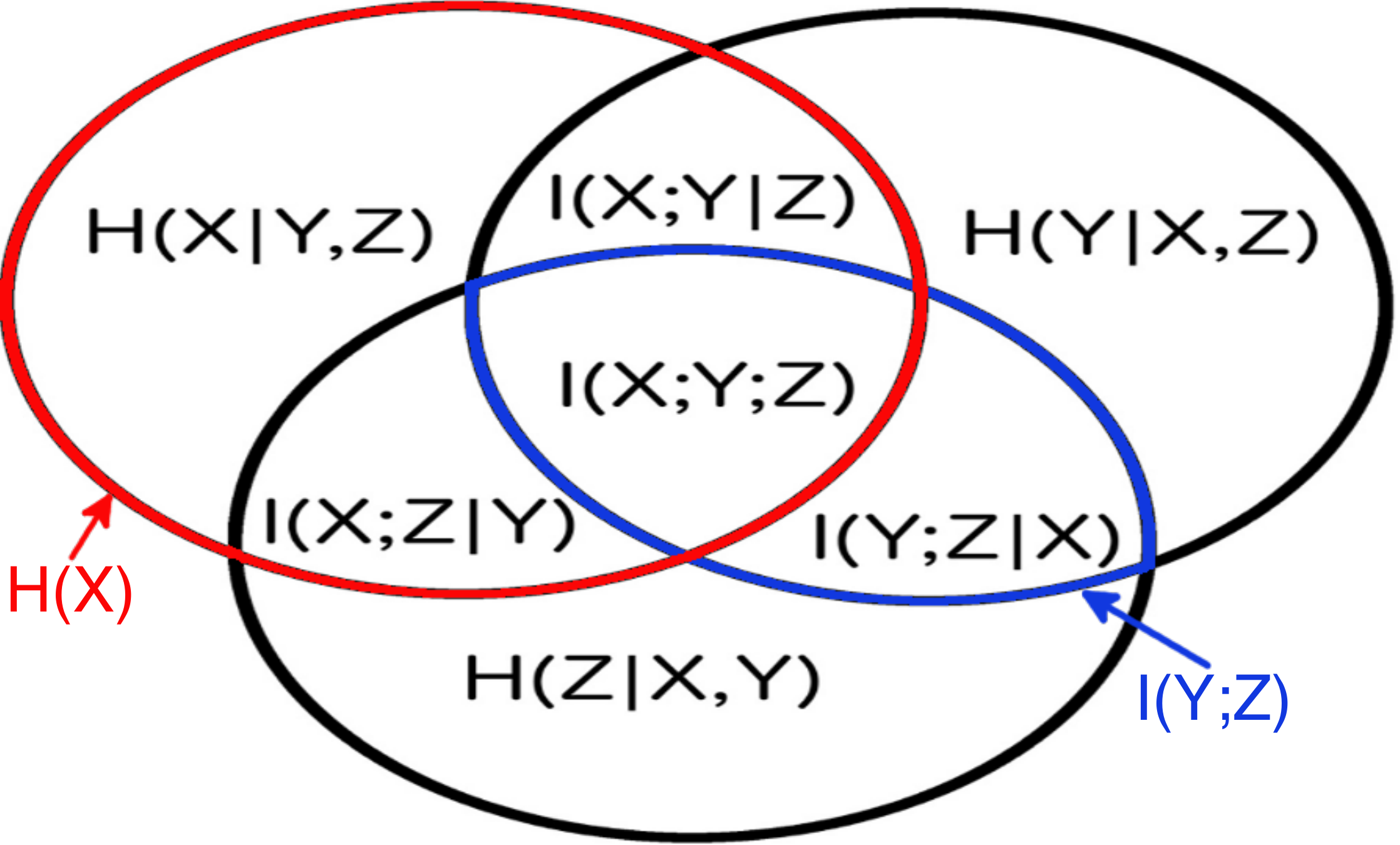}
   \end{center}
   \caption{\label{Fig:InfoDiag}(Color online) The information diagram
     for three variables. It contains 7 regions corresponding to the
     possible combinations of 3 variables, with their corresponding
     information-theoretic quantities defined in the text. The
     univariate entropy $H(X)$ is the sum of all the regions in the
     red circle, and the mutual information $I(Y;Z)$ is the sum of all
     the regions in the blue oval.}
\end{figure}

The set-theoretic formulation of information theory maps
information-theoretic quantities to regions of an information
diagram~\cite{Yeung2008information}, which is a variation of a Venn
diagram. The information diagram for three variables is shown in
Fig.~\ref{Fig:InfoDiag} with the associated information-theoretic
quantities labeled
\footnote{We use the convention $p(x,y,z) = P(X = x, Y = y, Z = z)$.}: 
entropy, $H(X) = \sum p(x) \log(p(x))$; conditional entropy, $H(X|Y,Z)
= \sum p(x,y,z) \log(p(x | y, z))$; mutual information, $I(X,Y) = \sum
p(x,y) \log(p(x,y)/(p(x)p(y)))$; conditional mutual information,
$I(X;Y|Z) = \sum p(x,y,z) \log \left(p(x;y|z)/[p(x|z)p(y|z) ]
\right)$; multivariate mutual information, $I(X;Y;Z) = I(X;Y) -
I(X;Y|Z)$.  In general the region where exactly $n$ variables
intersect corresponds to the $n$-variate mutual information between
those $n$ variables conditioned on the remaining variables.

%%%%%%% FIG %%%%%%%%%%%%%%%%%%%%%%%%%%%%%%%%%%%%%%%%%%%%%%%%%%%%%%%%%
%%%%% FIG %%%%%%%%%%%%%%%%%%%%%%%%%%%%%%%%%%%%%%%%%%%%%%%%%%%%%%%%%

% {\bf M:1: Many, most, information theoretic quantities can be written
%   as sums of atoms of the information diagram, ex. all entropies and
%   mutual informations on the set of variables; We use this fact to
%   solve for the maximum entropy given constraints of this type using
%   linear optimization; We also have a greedy algorithm that can solve
%   for this very rapidly}

Many information-theoretic quantities of interest can be written as a
sum of the `atoms' of information diagrams --- their smallest
subunits. For example, all entropies, mutual informations, and their
conditioned variants can be expressed in this way. Given constraints
of this form we can calculate the maximum entropy using linear
optimization.

% {\bf M:2: Our method work by constructing the information diagram
%   with the largest possible entropy given the constraints; intuitively
%   corresponds to maximally disjoint diagram, example conditioning on
%   the univariate entropies alone results in completely disjoint
%   diagram; things get much more complicated when conditioning on other
%   quantities such as mutual informations}

Our methods works by constructing the information diagram with the
largest entropy given the constraints, which intuitively corresponds
to creating the maximally disjoint diagram. For example, conditioning
on the univariate entropies alone results in the diagram being
completely disjoint, i.e., the maximum entropy is the sum of the
univariate entropies --- a well known result. However, when
conditioning on other terms, such as mutual informations, calculating
the maximum entropy is no longer straightforward.

Mutual informations and entropies correspond to regions of the
information diagram and can be written as the sum of the atoms in
their region.  In general, a system of $N$ variables, $\{X\}_N$, will
have $2^N-1$ atoms corresponding to all possible possible combinations
of the variables, excluding the empty set.  To illustrate this, for
the three variable case shown in Fig.~\ref{Fig:InfoDiag} we can see
the decompositions

\begin{eqnarray}
  \label{Eq:IConst}
  I(Y;Z) &=&  I(Y;Z|X) + I(X;Y;Z) \\
  \nonumber
  H(X) &=& H(X|Y,Z) + I(X;Y|Z) \\
  &&+ I(X;Z|Y) + I(X;Y;Z).
\end{eqnarray}

% {\bf M:?: In general any set of conditions which bound the maximum
%   entropy could be used; Illustration in Applications where we show
%   that }

Any set of information-theoretic quantities can be used as constraints
with our method --- as long as they can be written as a linear
function of the atoms of the information diagram and they bound the
total entropy.  This includes all $k$-variate conditional entropies
and $k$-variate conditional mutual informations. The $k$-variate
entropy of a set of variables $\{X\}_k$ conditioned on a set of
variables $\{X\}_n$ will be the sum of the atoms in the set $\{X\}_k$
excluding those also in $\{X\}_n$, e.g., $H(X,Y|Z) = H(X|Y,Z) +
H(Y|X,Z) + I(X;Y|Z)$, Fig.~\ref{Fig:InfoDiag}. Similarly, the
$k$-variate mutual information between a set of variables $\{X\}_k$
conditioned on a set of variables $\{X\}_n$ will be the sum of the
atoms that are in the intersection of all $k$ variables, but not in
any atoms corresponding to $\{X\}_n$. If these are all the variables
in the diagram this will be a single atom e.g., $I(X;Y|Z)$ in
Fig.~\ref{Fig:InfoDiag}. We illustrate this further in
Sec.~\ref{SSec:ConectInfo} where we condition on $n$-variate
entropies.

% {\bf M:?: Have additional Shannon inequalities which must be
%   satisfied; All Shannon inequalities can be constructed from
%   elemental inequalities of the form; This is a minimal set since none
%   of these inequalities are implied by any combination of the others;
%   Each of these inequalities can also be represented as sums over
%   atoms of the diagram}

In addition to any constraints one chooses, for discrete variables,
the information diagram must satisfy all the Shannon inequalities to
be valid, i.e. for there to exist a probability distribution with
those information-theoretic quantities.  All Shannon inequalities can
be constructed from elemental inequalities of the following two forms:

\begin{equation}
  \label{Eq:Hineq}
  H(X_i|\{X\}_N - X_i) \geq 0
\end{equation}

\noindent
and

\begin{equation}
  \label{Eq:Iineq}
  I(X_i,X_j| \{X\}_K) \geq 0, 
\end{equation} 

\noindent
where $i \neq j$, and $\{X\}_K \subseteq \{X\}_N - \{X_i,X_j\}$. For
continuous random variables entropies can be negative, so inequalities
of the form Eq.~(\ref{Eq:Hineq}) are not applicable, though those of
the form Eq.~(\ref{Eq:Iineq}) still are. This is a minimal set of
inequalities as no inequality is implied by a combination of the
others. Each of these inequalities can also be written as the sum of
atoms in their region. This is trivial for inequalities like
Eq.~(\ref{Eq:Hineq}) since all $H(X_i | \{X\}_N - X_i)$ are themselves
atoms. There will also be ${N \choose 2}$ inequalities like
Eq.~(\ref{Eq:Iineq}) that are atoms of the diagram. For four variables
a nontrivial decomposition into atoms of an Eq.~(\ref{Eq:Iineq})
inequality is

\begin{equation}
  \label{Eq:ShannonIneq}
  I(X_1;X_2|X_3) = I(X_1;X_2|X_3,X_4) + I(X_1;X_2;X_4|X_3) \geq 0.
\end{equation}

% {\bf M:5: There also exists non-Shannon inequalities which we don't
%   account for; While our diagram will satisfy the given linear
%   equalities and Shannon inequalities it may violate the non-Shannon
%   inequalities; Our diagram may not be realizable but it will still
%   represent an upper limit on the possible entropy}

There also exists so called non-Shannon inequalities for $N\geq 4$,
which are not deducible from the Shannon
inequalities~\cite{Yeung2008information}. While it is possible, in
principle, to include these in our maximization they have not yet been
fully enumerated. Therefore, we restrict the set of inequalities we
use to the Shannon inequalities. As the diagram may violate a
non-Shannon equality, there may be no probability distribution that
satisfies it.  However, the diagram would still represent an upper
bound on the possible entropy.

% {\bf M:6: For a large class of diagrams we do know that they can be
%   realized; When all the regions of the diagram are positive we know
%   that there will be an infinite number of distributions which satisfy
%   it}

For a large class of diagrams we do know our bound is achievable. We
prove in Sec.~\ref{SSec:ConstructPositivity} that whenever all the
atoms of the diagram are non-negative it is possible to construct a
set of variables that satisfy it.  It is easy to see from this proof
that there will in fact be an infinite number of distributions
satisfying the diagram in these cases. There will of course also be
many diagrams with negative regions that are also satisfiable, but our
constructive proof can not verify this.

We have now shown that the task of finding the maximum entropy,
conditioned on the information-theoretic quantities discussed here, as
well as the elemental Shannon inequalities, can be solved using linear
optimization.  Each constraint will take the form of a linear equality
or inequality, as in Eq.~(\ref{Eq:IConst}) and~(\ref{Eq:ShannonIneq}),
and we maximize the N-variate entropy by maximizing the sum over all
$A$ atoms of the information diagram.

% {\bf M:7: Our method finds the maximum entropy over arbitrary
%   cardinality variables; Small cardinality variables may not be able
%   to achieve our maximum. }

Our method is free of distributional assumptions, finding the maximum
entropy possible for variables of any cardinality given only
information-theoretic constraints. This can result in the maximum
entropy diagram being unconstructable for low cardinality variables,
even though it is achievable for higher cardinality ones. However this
does not seem to be a large issue in practice, as can be seen in our
results in~\cite{Martin2015::PRL}.

% {\bf M:8: Usining linear optimization we could also find the minimum
%   possible entropy given the constraints; Constructive proof of
%   existance is much less likely to hold since there is a much higher
%   chance of regions being negative; we focus on the maximization
%   however because of it's relevance to statistical inference problems}

Given information-theoretic constraints of the type we have been
discussing, it is just as easy to use linear optimization to find the
minimum possible entropy as it is to find the maximum. The minimum
entropy diagram is much more likely to have negative regions though,
so our constructive proof of existence is not likely to hold in these
cases. Analogous to the maximum entropy diagram, the minimum diagram
will still represent a lower bound on the possible entropy. We focus
on the maximum case because of its use in statistical physics, and
more generally in statistical inference.

\section{Proofs}
\label{Sec:Proofs}

\subsection{If an information diagram has only non-negative regions it
  can always be constructed}
\label{SSec:ConstructPositivity}

Given an information diagram for a set of $N$ variables, $\{X\}_N$,
with atoms $\{A\}$, and all $A_j = a_j \geq 0$, we can always
construct a probability distribution of $N$ variables that would have
this diagram.  We introduce a set of variables $\{Y\}$ which we define
to be independent and have entropies $H(Y_j) = a_j$; every region $A_j
= a_j$ is associated with an independent random variable with entropy
$a_j$. Each variable $X_i$ is now defined to be the set of $Y_j$ that
have regions which lie in $H(X_i)$.

The set of variables, $\{X\}_N$, will satisfy all the information
regions of the diagram. We will prove this by showing that $\{X\}_N$
will reproduce all $H(\{X\}_n)$, where $\{X\}_n$ is an $n$-variate
subset of $\{X\}_N$. All the regions of the information diagram can be
calculated from the set of $H(\{X\}_n)$, so if $\{X\}_N$ reproduces this
set it will reproduce the entire diagram.

The set $\{X\}_n$ will be the set of all $Y_j$ with a region
associated with any of the $n$ variables in $\{X\}_n$. Of course some
$Y_j$ will be included more than once, but this will not affect the
entropy since $H(Y_j,Y_j,\{Y\}_l) = H(Y_j,\{Y\}_l)$.  The entropy
$H(\{X\}_n)$ would then be the sum of all the associated entropies
$H(Y_j)$, since all $Y_j$ are independent by definition. The sum of
the entropies of $H(Y_j) = a_j$ is the same as the sum of all the
regions in the information diagram associated with $H(\{X\}_n)$, and
hence $\{X\}_n$ will satisfy all such entropies.

\subsection{Analytical Maximum for $N=3$}
\label{SSec:ProofN3}

When conditioning on bivariate mutual informations and univariate
entropies we have an analytical solution for the maximum entropy when
$N=3$. For three variables we can write the joint entropy as

\begin{equation}
  \label{Eq:3VarEnt}
  H = \sum_iH(X_i) - \sum_{i>j} I(X_i;X_j) + I(X_1;X_2;X_3).
\end{equation}

\noindent
We can see why Eq.~(\ref{Eq:3VarEnt}) is true by imagining the
information diagram, and realizing the total entropy must be the sum
of all its elements. By adding all the univariate entropies all the
conditional entropies in the information diagram are added once, but
all the regions of overlap are added multiple times. These multiple
counts are then removed when we remove all the mutual informations,
but now we remove regions where more then 2 variables overlap too many
times. For three variables we then need to add back the triplet region
once. It was added three times by the entropies and removed three
times by the mutual informations.

Since we are conditioning on the univariate entropies and mutual
informations, the only free parameter is

\begin{equation}
  \label{Eq:3VarMI}
  I(X_1;X_2;X_3) = I(X_1;X_2) - I(X_1;X_2|X_3).
\end{equation}

\noindent
This means that the maximum of Eq.~(\ref{Eq:3VarEnt}) will occur when
Eq.~(\ref{Eq:3VarMI}) is maximal. Both $I(X_1;X_2)$ and
$I(X_1;X_2|X_3)$ must be positive, so Eq.~(\ref{Eq:3VarMI}) can be no
greater than the minimum mutual information between $X_1$, $X_2$, and
$X_3$.

\begin{figure}[!h]
%%\begin{figure}[htbp]
  \begin{center}
    \includegraphics*[width=\columnwidth]{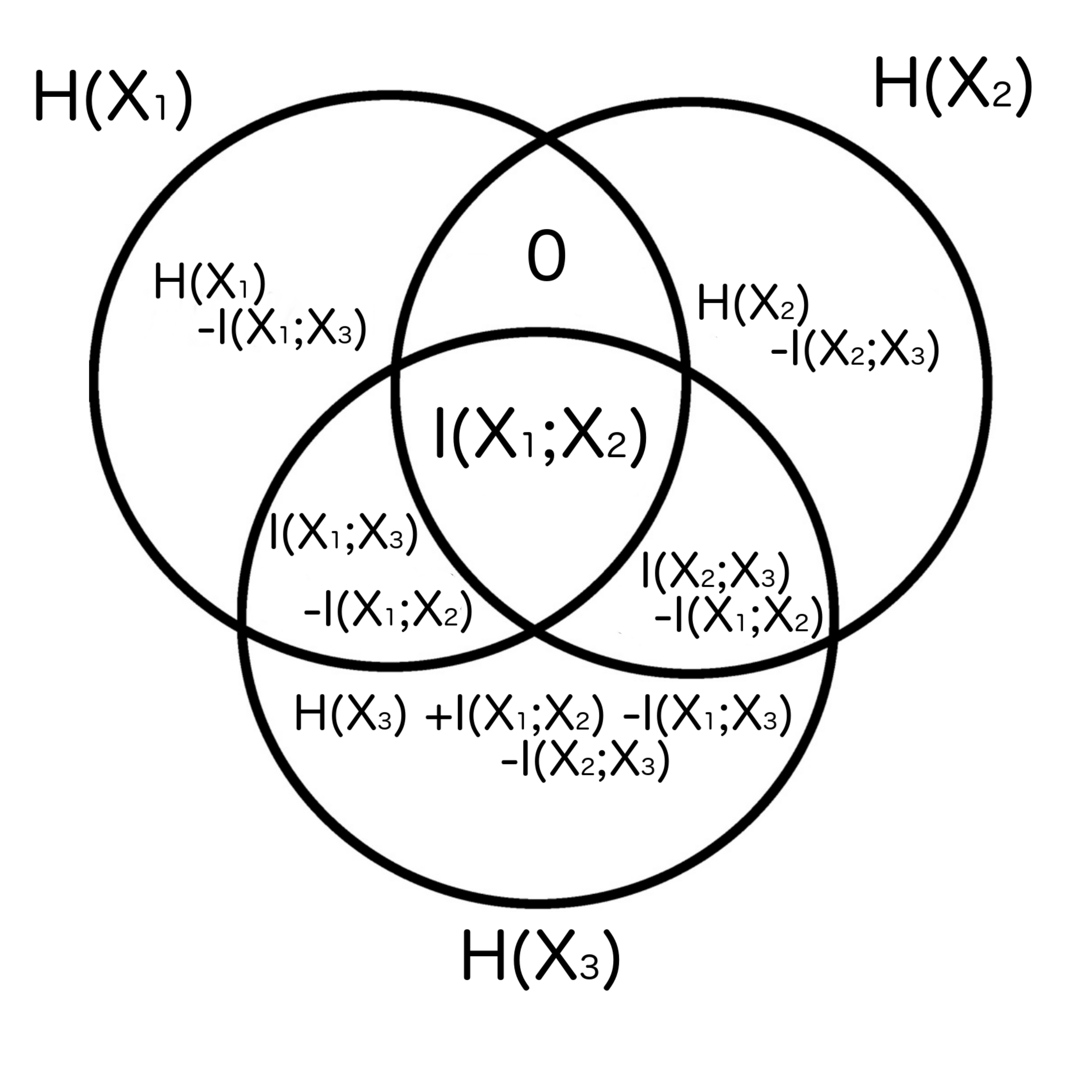}
   \end{center}
   \caption{\label{Fig:3VarMaxDiag}The maximum entropy diagram for
     three variables if the minimum mutual information between the
     variables is $I(X_1;X_2)$.}
\end{figure}

We now show that we can always construct this diagram when the
variables are discrete since it will only have non-negative regions.
Without loss of generality we can define the minimal mutual
information to be $I(X_1;X_2)$. This results in the information
diagram in Fig.~\ref{Fig:3VarMaxDiag}. By inspection we can see that
this diagram satisfies the constraints on the univariate entropies and
mutual informations. Since $I(X_1;X_2)$ is the minimal mutual
information, and all the mutual informations are non-negative, all the
regions where multiple variables overlap in the diagram are
non-negative. Now we must show that all the conditional entropies in
the diagram are non-negative. The mutual information between two
discrete variables can not be greater than their univariate entropies,
therefore $H(X_1|X_2,X_3) \geq 0$ and $H(X_2|X_1,X_3) \geq 0$.

%%%%%%% FIG %%%%%%%%%%%%%%%%%%%%%%%%%%%%%%%%%%%%%%%%%%%%%%%%%%%%%%%%%
%%%%% FIG %%%%%%%%%%%%%%%%%%%%%%%%%%%%%%%%%%%%%%%%%%%%%%%%%%%%%%%%%

The final part now is to prove that $H(X_3|X_1,X_2) \geq 0$, which we
show is true provided that the constraints are satisfiable. We now
look solely at the regions inside $H(X_3)$, and look at the affect of
adding $\epsilon$ to the $I(X_1;X_2;X_3)$ region. To conserve the
univariate entropy and mutual informations associated with $X_3$, we
must make the following changes

\begin{eqnarray}
  I(X_1;X_3|X_2) \rightarrow I(X_1;X_3|X_2) - \epsilon \\
  I(X_2;X_3|X_1) \rightarrow I(X_2;X_3|X_1) - \epsilon \\
  H(X_3|X_1,X_2) \rightarrow H(X_3|X_1,X_2) + \epsilon.
\end{eqnarray}

\noindent
We see from this that changing one region in $H(X_3)$ necessitates
changing all the regions in $H(X_3)$. We also see that changing
$I(X_1;X_2;X_3)$ changes $H(X_3|X_1,X_2)$ by the same amount. This
means that the largest $H(X_3|X_1,X_2)$ can be is when
$I(X_1;X_2;X_3)$ is also maximal -- as in our maximal construction,
Fig.~\ref{Fig:3VarMaxDiag}. Therefore if our constructed case resulted
in $H(X_3|X_1,X_2) < 0$ the constraints are unsatisfiable since this
is the largest that $H(X_3|X_1,X_2)$ can be made.

Figure~\ref{Fig:3VarMaxDiag} shows that the maximum entropy,
conditioned on bivariate mutual informations and univariate entropies,
corresponds to the pair of variables with the smallest mutual
information being conditionally independent. This is notable, as it is
essentially what is done in~\cite{Margolin2006::BMCbioinfo}, where
they attempt to infer interactions between genes; for every triplet of
genes they consider the pair with the smallest mutual information to
be independent. While they justify this using the data processing
inequality~\cite{Cover2006}, our proof here lends this procedure
further credibility.

\subsection{Proof that conditioning on the first two moments is
  equivalent to conditioning on bivariate distributions for binary
  variables}

Maximizing the joint entropy of a set of binary variables, conditioned
on their first two moments, is the same as conditioning on the joint
probability distributions. The univariate distributions can be
reconstructed from the first moments

\begin{eqnarray}
  E[X] &=& x_0p(x_0) + x_1(1-p(x_0)) \\
  p(x_0) &=& \frac{E[X] - x_1} {x_0 - x_1}.
\end{eqnarray}

\noindent
This information plus the covariances exactly specify the bivariate
distributions. For the bivariate distributions we have

\begin{eqnarray}
  p(x_0,y_0) + p(x_1,y_0) = p(y_0)\\  
  p(x_0|y_0)p(y_0) + p(x_1|y_0)p(y_0) = p(y_0)\\  
  p(x_0|y_0) + p(x_1|y_0) = 1 \\
  p(x_0|y_0) + \frac{p(x_1) - p(x_1|y_1)p(y_1)}{p(y_0)} = 1\\
  p(x_0|y_1) + p(x_1|y_1) = 1   
\end{eqnarray}

Therefore, for the 2-variable conditional probabilities there is only
one degree of freedom when the marginal probabilities are known, which
is equivalent to the covariance

\begin{eqnarray}
  \nonumber
  C[X,Y] &=& x_0y_0p(x_0,y_0) + x_0y_1p(x_0,y_1) + x_1y_0p(x_1,y_0)\\ \nonumber
  && + x_1y_1p(x_1,y_1)\\
  \nonumber
  p(x_0|y_0) &=&  [C[X,Y] -x_0y_1p(x_0) - x_1y_0p(y_0) \\ 
  \nonumber
  && +x_1y_1(p(y_0) -p(x_1)) ] \\
  \nonumber
  && \times \left[ p(y_0) (x_0y_0 - x_1y_0 -x_0y_1 + x_1y_1 ) \right]^{-1}.
\end{eqnarray}

\noindent
Therefore, maximizing the entropy conditioned on the first two moments
of a set of binary variables is equivalent to maximizing the entropy
conditioned on their bivariate probability distributions.

\section{Applications}
\label{Sec:App}

\subsection{Undersampled Regime}
\label{SSec:UnderSamp}

Possibly one of the most exciting applications of our method is in the
undersampled regime. It is possible to estimate the entropy of a set
of discrete variables with $n \sim 2^{H/2}$ samples (where $H$ is
measured in bits)~\cite{Nemenman2011::Ent}. This means it is possible
to make maximum entropy estimates even when the marginal probability
distributions have not been sufficiently sampled, as needed to
calculate the connected informations in~\cite{Schneidman2003::PRL}.

As an example, consider an Ising type model with probability distribution,

\begin{equation}
\label{Eq:Ising}
P(\mathbf{X}) = \frac{1}{Z} \exp \left( \sum_{i=1}^N h_i x_i +\sum_{i > j} J_{i,j} x_i x_j \right),
\end{equation}

\noindent
where $Z$ is a normalization constant. These distributions often
arises in the context of establishing the importance of pairwise
interactions, because they describe the maximum entropy distribution
consistent with the first two moments of a set of
variables~\cite{Martin2015::PRL,Timme2014}. Therefore, we would expect
the difference between the entropy of the true distribution and the
maximum entropy conditioned on the bivariate distributions to be zero.

\begin{figure}[!h]
%%\begin{figure}[htbp]
  \begin{center}
    \includegraphics*[width=\columnwidth]{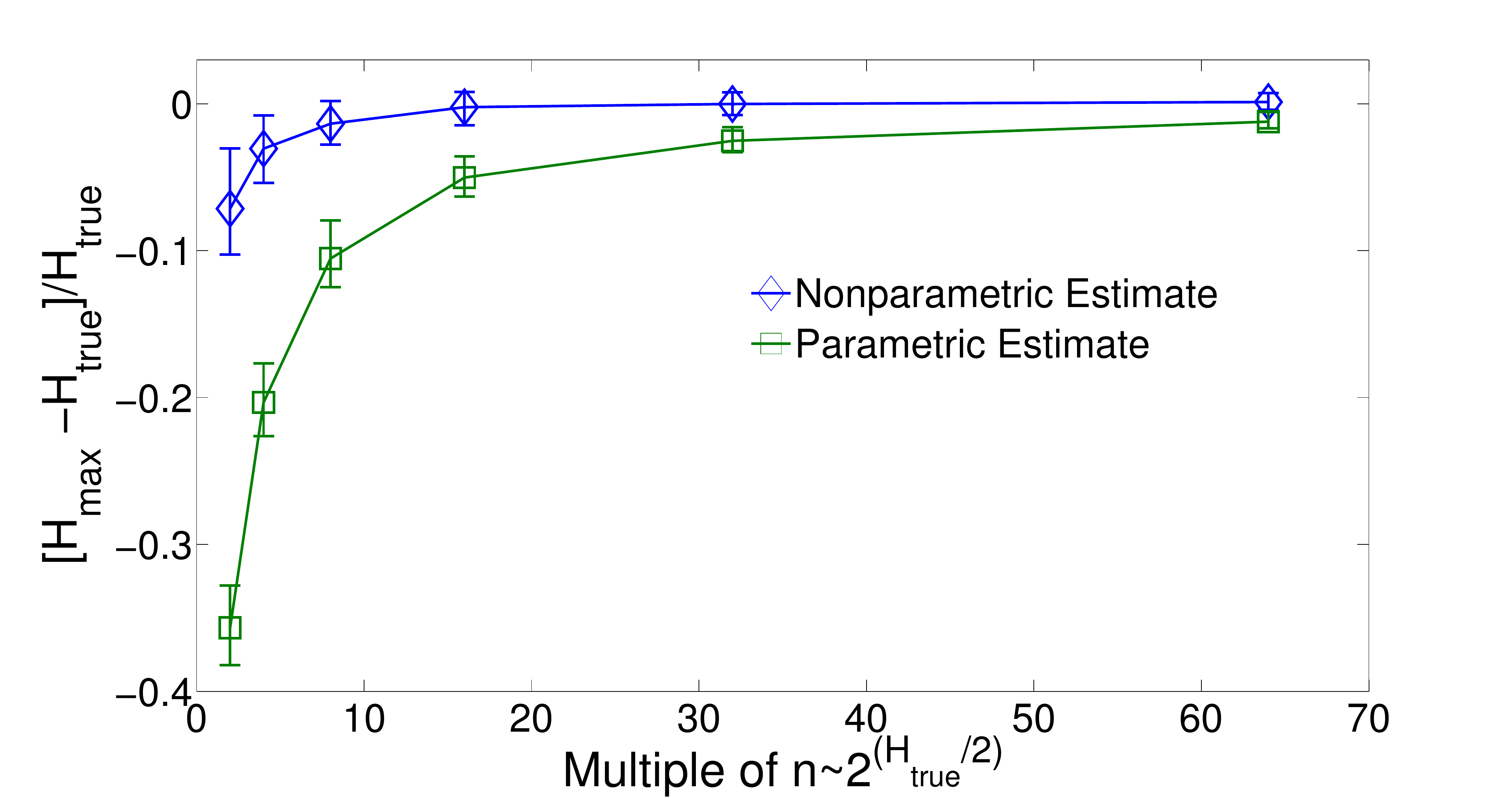}
   \end{center}
   \caption{\label{Fig:UnderSamp}(Color online) The fractional
     difference between the maximum entropy estimate and the true
     entropy using our nonparametric maximum calculated from the
     univariate and bivariate entropies, as well as estimating the
     maximum parametrically from the estimated bivariate probability
     distributions. One hundred distributions of three variables of
     the form Eq.~(\ref{Eq:Ising}) were generated with the parameters
     $h_i$ and $J_{i,j}$ drawn from normal distributions with zero
     mean and standard deviation $0.1$, with each variable having a
     cardinality of 5. The minimum number of samples needed, $n \sim
     2^{H/2}$, ranged from 4 to 12. The error bars are given by the
     $25\%$ and $75\%$ quantiles.}
\end{figure}

At small sample sizes however, the maximum entropy is severely
underestimated when conditioning on naively estimated bivariate
distributions. On the other hand, a much more accurate estimate of the
maximum entropy is obtained when estimating the univariate and
bivariate entropies using the estimator in~\cite{Nemenman2011::Ent},
and using these as constraints in our nonparametric method.  This is
shown in Fig.~\ref{Fig:UnderSamp}.

%%%%%%% FIG %%%%%%%%%%%%%%%%%%%%%%%%%%%%%%%%%%%%%%%%%%%%%%%%%%%%%%%%%
%%%%% FIG %%%%%%%%%%%%%%%%%%%%%%%%%%%%%%%%%%%%%%%%%%%%%%%%%%%%%%%%%

% \subsection{Identifying Nonlinear Relationships}

% {\bf The diagrammatic and moment methods can be used together to
%   identify the existance of nonlinear relationships in the system; If
%   the diagrammatic method provides a much tighter bound on the entropy
%   than the momment method this means that there must be significant
%   nonlinear relationships in order to account for this discrepency}

\subsection{Computation Time}
\label{SSec:ComputTime}

To illustrate the potential computational speedups possible using our
methods, we consider Ising type distributions, Eq.~(\ref{Eq:Ising}),
again. Specifically, we investigate the dependence on different
numbers of random variables, $N$, and variable cardinality. In each
case the parameters $h_i$ and $J_{i,j}$ are drawn from a normal
distribution with mean zero and variance $0.1$.

\begin{figure}[!h]
%%\begin{figure}[htbp]
  \begin{center}
    \includegraphics*[width=\columnwidth]{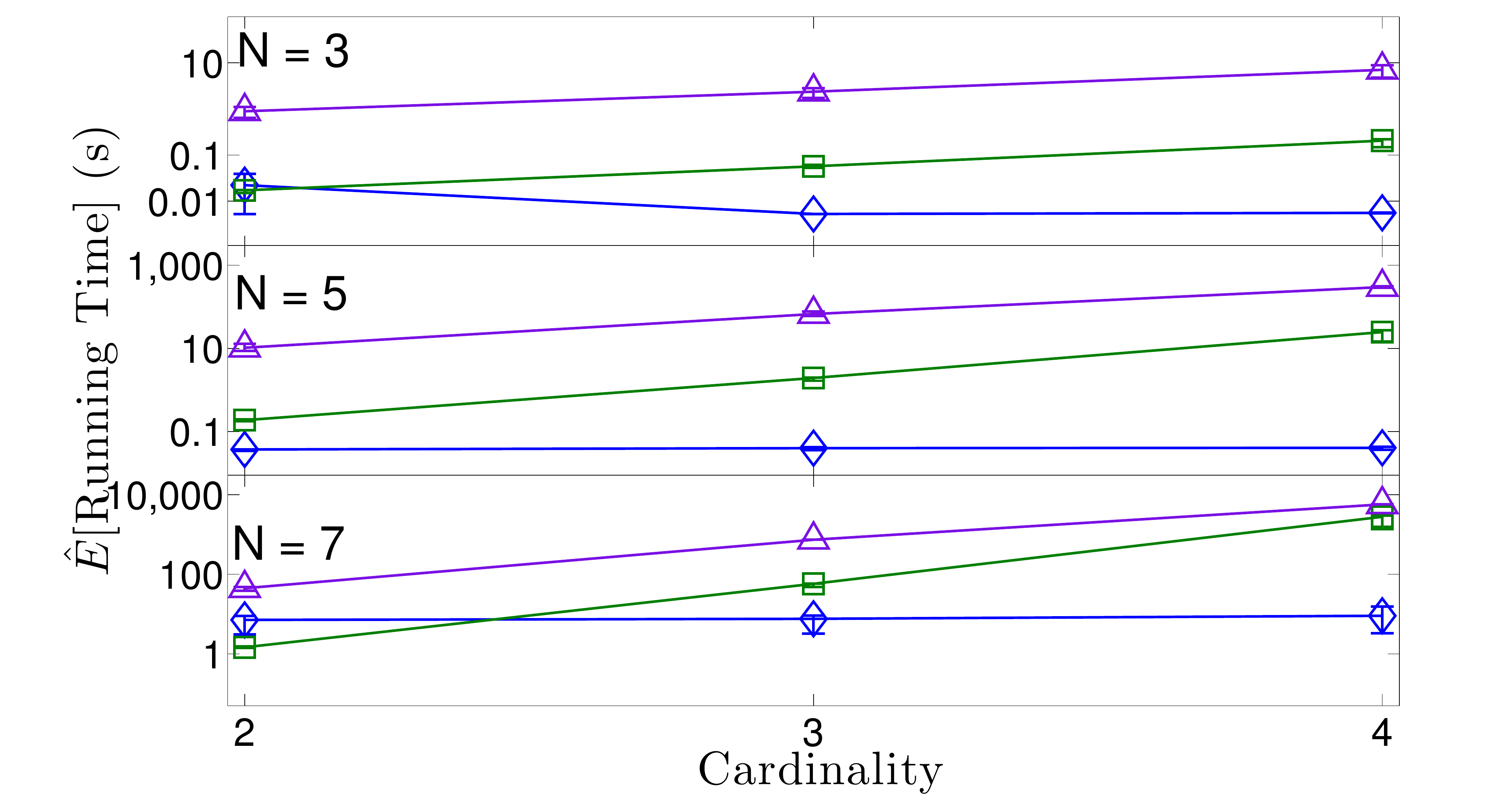}
   \end{center}
   \caption{\label{Fig:TimeTrial} (Color online) The expected running
     time for different methods at different variable cardinalities,
     and number of variables, $N$. The three methods are: our linear
     optimization method conditioned on mutual informations and
     univariate entropies (blue diamonds); iterative fitting
     conditioned on the bivariate distributions (green squares);
     iterative fitting conditioned on the first two moments (purple
     triangles). For each cardinality and $N$, the distributions and
     error calculations are the same as for Fig.~\ref{Fig:UnderSamp}.}
\end{figure}

Figure~\ref{Fig:TimeTrial} compares the runtime of our algorithm with
that using iterative proportional
fitting~\cite{Darroch1972::AnMatStat}, where we show both conditioning
on the bivariate distributions and conditioning on the first two
moments of the distributions. Since our method only uses
information-theoretic quantities as inputs it is not affected by the
cardinality of the variables, i.e., if the variables have a
cardinality of two or 100 it will have no bearing on how long our
method takes to run, as long as the information-theoretic quantities
conditioned on are the same. As the other methods do depend on the
cardinality of the variables we expect that at `some' cardinality our
method will certainly outperform them. In fact, as
Fig.~\ref{Fig:TimeTrial} shows, only when the variables have a
cardinality of two are the runtimes comparable, with our method
running orders of magnitude faster at all measured higher
cardinalities.

%%%%%%% FIG %%%%%%%%%%%%%%%%%%%%%%%%%%%%%%%%%%%%%%%%%%%%%%%%%%%%%%%%%
%%%%% FIG %%%%%%%%%%%%%%%%%%%%%%%%%%%%%%%%%%%%%%%%%%%%%%%%%%%%%%%%%

\subsection{Estimating connected informations}
\label{SSec:ConectInfo}

Next we show how our method can be used to nonparametrically estimate
connected informations~\cite{Schneidman2003::PRL}, which are useful
for estimating the relevance of higher-order interactions in sets of
variables. However, as the number of variables and their cardinality
increases, estimating these values from the probabilities directly can
suffer from lack of samples do to the exponentially increasing phase
space, as well as quickly become computationally intractable.

In order to estimate connected informations using our method we
condition on $n$-variate entropies. This lets us answer the question
of what the maximum possible $N$-variate entropy is given any set of
entropies. Let $a_{\{X\}_k}$ be the set of all atoms that lie in the
joint entropy $H(\{X\}_k)$. The univariate entropy $H(X_i)$ is the sum
of all atoms $a_{X_i}$.  Similarly, the bivariate entropy $H(X_i,X_j)$
is the sum of all atoms $a_{\{X_i,X_j\}}$. This is easily generalized
to the $n$-variate entropy $H(\{X\}_n)$, which is the sum of all atoms
$a_{\{X\}_n}$. Therefore, we can use any $n$-variate entropy as a
constraint in the linear optimization problem.

The connected information of order $k$ is,

\begin{equation}
  \label{Eq:ConnectSch}
  I_C^{(k)}(\{X\}_N) = H[\widetilde{P}^{(k-1)}(\{X\}_N)] -  H[\widetilde{P}^{(k)}(\{X\}_N)],
\end{equation}

\noindent
where $\widetilde{P}^{(k)}(\{X\}_N)$ is the maximum entropy
distribution consistent with all $k$-variate marginal
distributions. Instead of this we propose an alternate expression

\begin{equation}
  \label{Eq:ConnectMe}
  I_C^{(k)} = \widetilde{H}^{(k-1)} -  \widetilde{H}^{(k)},
\end{equation}

\noindent
where $\widetilde{H}^{(k)}$ is the maximum entropy consistent with all
the one through $k$-variate entropies. 

This formulation has three advantages: 1) estimating the $k$-variate
marginal distributions can be problematic do to insufficient data,
whereas much better estimates of the $k$-variate entropies may be
available such as~\cite{Nemenman2011::Ent}, as we showed in
Sec.~\ref{SSec:UnderSamp}; 2) this equation is easily estimated using
our maximum entropy estimation, which can offer significant
computational speedups over existing techniques, as we showed in
Sec.~\ref{SSec:ComputTime}; 3) this can be estimated given just the
information-theoretic quantities independent of any specific knowledge
of the underlying distributions.

It is important to realize though that Eqs.~(\ref{Eq:ConnectSch})
and~(\ref{Eq:ConnectMe}) differ in that the latter does not constrain
the cardinality of the variables and could violate the non-Shannon
inequalities as discussed in Section~\ref{Sec:Method}. Therefore, it
will always be the case that $\widetilde{H}^{(k-1)} \geq
H[\widetilde{P}^{(k-1)}(\{X\}_N)]$. In the examples we have looked at
before~\cite{Martin2015::PRL}, as well as in the illustrative example
we give next, this does not seem to appreciably affect the results
however.

\subsubsection{Illustrative Example}
\label{SSSec:Examp}

The quintessential example of an entirely 3-variate interaction is the
Exclusive OR (XOR) gate when the inputs are chosen uniformly and
independently, the truth table of which is given in
Table~\ref{Tab:XOR}.  Any pair of variables taken alone appear to be
independent, though given the state of two the state of the third is
uniquely determined. This can be generalized to an $N$-variate
relationship by taking $N-1$ independently generated random variables
uniformly drawn from the set $\{0,1\}$, and the $N$th their sum modulo
two. We can also generalize to arbitrary cardinalities, $C$, by
drawing the $N-1$ variables independently and uniformly from the set
$\{0,..,C-1\}$, and the $N$th is now their sum modulo $C$.

\begingroup
\squeezetable
\begin{table}
  \begin{tabular}{l c | c}
    $X$ & $Y$ & $Z$ \\ \hline
    0 & 0 & 0 \\
    1 & 0 & 1 \\
    0 & 1 & 1 \\
    1 & 1 & 0.
  \end{tabular}
  \caption{\label{Tab:XOR}Truth table for an Exclusive OR (XOR) gate, where the inputs are
    $X$ and  $Y$, and the output is $Z$.}
\end{table}
\endgroup

\vspace{0.1mm}

We now show that in these cases our nonparametric connected
information, Eq.~(\ref{Eq:ConnectMe}), will return the same result as
the parametric one, Eq.~(\ref{Eq:ConnectSch}). Given a set of $N$
variables with cardinality $C$, and an $N$-variate interaction of the
type discussed above, the joint entropy of any set of $k<N$ variables
will be the sum of the univariate entropies, $H(\{X\}_k) = \sum
H(X_i)$. This means for both Eq.~(\ref{Eq:ConnectMe})
and~\ref{Eq:ConnectSch}, $I_c^{(k)} = 0$ for $k<N$. For $k=N$, both
$\widetilde{H}^{(k)}$ and $H[\widetilde{P}^{(k)}(\{X\}_N)]$ are the
true $N$-variate entropies, and $I_c^{(k)} = H(X_i)$ in both cases. We
can see from this that both methods will also return the same result
for a system of $N$ variables that is composed of independent sets of
$n$ variables with $n$-variate relationships, where $n$ is allowed to
differ between sets, e.g. two XOR gates, where $N=6$ and $n=3$ for
both sets.

\subsection{Resting-State Human Brain Networks}
\label{SSec:RestNet}

To illustrate the applicability of the described methodology in
real-world data situations, we apply it to neuroimaging data, in a
similar context as in the recent study by Watanabe et
al~\cite{Watanabe2013::NatCom}. In particular, we want to assess to
what extent the multivariate activity distribution is determined by
purely bivariate dependence patterns. This is of relevance because the
use of bivariate dependence matrices, particularly of pairwise
correlations, is currently a prominent method of characterizing the
brain interaction structure.  If pairwise relationships are sufficient
to describe the interaction structure of the brain this would
tremendously simplify the task of uncovering this structure. If this
were not the case, it would mean that higher-order relationships, as
discussed in Sec.~\ref{SSec:ConectInfo}, would need to be analyzed. As
the phase space of the problem grows exponentially as we probe ever
higher-order interactions, this would result in us rapidly running out
of sufficient data to sample these spaces, and measure the
corresponding interactions.

The used data consist of time series of functional magnetic resonance
imaging signal from 96 healthy volunteers measured using a 3T Siemens
Magnetom Trio scanner in IKEM (Institute for Clinical and Experimental
Medicine) in Prague, Czech Republic. Average signals from 9 regions of
the well-known default mode network, and 12 regions of the
fronto-parietal network were extracted using a brain
atlas~\cite{Shirer2011}. Standard preprocessing and data denoising was
carried out using processing steps described in a previous
neuroimaging study~\cite{Hlinka2011Neuroimage}. The data were
temporally concatenated across subjects to provide a sufficient sample
of $T=36480$ timepoints. Each variable was further discretized to 2 or
3 levels using equiquantal binning. Entropies were then estimated
using the estimator in~\cite{Nemenman2011::Ent}. We tested that we
could estimate the full joint entropy by estimating it for increasing
sample sizes, and checking that the estimate stabilized for the largest
available sample sizes. Moving to larger cardinalities was not
possible due to insufficient data available to estimate the full joint
entropy of the resting-state networks.

Our analysis of the default mode network resulted in $I_m/I_N=1$ and
$0.90$ for the 2-level and 3-level discretizations respectively, when
conditioning on the first two moments, and $0.86$ and $0.90$ when
using our technique conditioned on bivariate mutual informations and
univariate entropies. Similarly, for the fronto-parietal network,
conditioning on the moments resulted in $I_m/I_N=1$ and $0.73$ for the
2-level and 3-level discretizations, and $0.77$ for both
discretizations when using our method. In both cases we can see that
conditioning on the first two moments resulted in a substantial
decrease in $I_m/I_N$ as the discretization was increased, while the
results using our method appear stable to the discretization. The
effect of discretization on both these methods is in accord with the
results for nonlinear model systems in~\cite{Martin2015::PRL}.

Overall, our findings are consistent with the observations reported
in~\cite{Watanabe2013::NatCom} for the 2-level and 3-level
discretization of the default mode network and the fronto-parietal
network, where they conditioned on the first two moments only. For
2-level discretization, they found $I_m/I_N=0.85$ and $0.96$ for the
default mode and fronto-parietal networks respectively. For the
3-level discretization, the ratio dropped to $I_m/I_N\approx0.55$ for
both networks. The variation between their specific values and ours
--- especially for the 3-level discretization --- is likely a result
of the different regions used to represent both networks in
combination with statistical variations starting from different data
sets to begin with. In conclusion, both their and our findings
indicate that nonlinear relationships play an important role in the
structure of fMRI data.

\subsection{Network inference}
\label{SSec:NetInf}

In the process of finding a maximum entropy estimate, the linear
optimization computes all the atoms of the information diagram. This
includes the conditional mutual information (CMI) between every
variable pair, conditioned on every other variable. This can be
interpreted as the level of direct pairwise interaction between
components of a dynamical system and thus be used as a novel method
for inferring structural connectivity, provided that the interactions
are predominately pairwise in the first place. In the following
section we show how using our entropy maximization conditioned on
mutual informations and univariate entropies outperforms other
techniques; in particular relevance networks as defined
in~\cite{butte2000mutual}, which work by picking a threshold for a
statistical similarity measure (in our case the mutual information),
and interpreting every pair of variables that cross this threshold as
directly interacting. To benchmark our method's performance we analyze
the Kuramoto model as a paradigmatic dynamical system with non-linear
coupling.

\subsubsection{The model}

The Kuramoto model was introduced in~\cite{kuramoto1975self} and consists 
of $N$ phase oscillators that are coupled in a particular topology. The 
$i$th oscillator's frequency is given by $\theta_i$ and its dynamics are 
described by

\begin{equation}
  \label{Eq:Kuramoto}
  \frac{\partial \theta _i}{\partial t} = \omega _i +\frac{K}{N} \sum_{j=1}^{N}\sigma_{ij} \sin (\theta_j-\theta_i) +\eta_i(t).
\end{equation}

Here $\omega _i$ is the natural frequency of the oscillator, and
$\eta_i(t)$ a random noise term drawn from a Gaussian distribution
with correlation function
$<\eta_i(t),\eta_j(t')>=G\delta_{ij}\delta(t-t')$, where G determines
the amplitude of the noise. $K$ represents a uniform coupling strength
between interacting nodes, and $\sigma_{ij} \in \{0,1\}$ represents
the adjacency matrix of the network, where $\sigma_{i,j} = 1$ for
connected nodes. The interactions are always taken to be
bidirectional, i.e. $\sigma_{ij}=\sigma_{ji}$. In the following, we
focus on the case when the adjacency matrix is an Erd\"os-R\`enyi
random graph~\cite{Erdos1960} of density $p$, with a fixed number of
links. The inference problem is then to reconstruct
$\boldsymbol{\sigma}$ from the measured time series $\theta_i$.

The time series are generated using the Euler-Maruyama method with a
step size $dt=2^{-6}$ and noise amplitude $G=0.05$.  The data gets
resampled such that only every 8th time step is used, and a transient
of $T_{trans}=50$ is removed.  Unless stated otherwise the network
size is $N=12$, the integration time $T=50,000$, the coupling strength
$K=0.5$, the number of links in the network $12$ (which corresponds to
each node having an average of 2 neighbors and $p\approx 0.18$).

The data is discretized using equiquantal binning into $n=3$
states. Numerical tests (using $n=5$ and $n=7$) have indicated that
larger cardinalities can improve the performance, given that the used
time series is long enough. Otherwise sampling issues may arise
(i.e. empty or almost empty bins) and degrade the quality of the
entropy estimation.  The intrinsic frequencies are drawn from a
uniform distribution on the interval $[\Omega, 3 \Omega]$ with $\Omega
= 20\cdot \frac{p}{N}$.  For higher values of $p$ synchronization
effects would be expected at lower coupling strengths. To counteract
this, the frequency scale increases with $p$. The distribution is
shifted away from zero to sample through the phase space more quickly,
i.e. avoid oscillators that stay in just one bin throughout the
system's time evolution.

\subsubsection{The method}

\begin{figure}[!h]
%%\begin{figure}[htbp]
  \begin{center}
    \includegraphics*[width=\columnwidth]{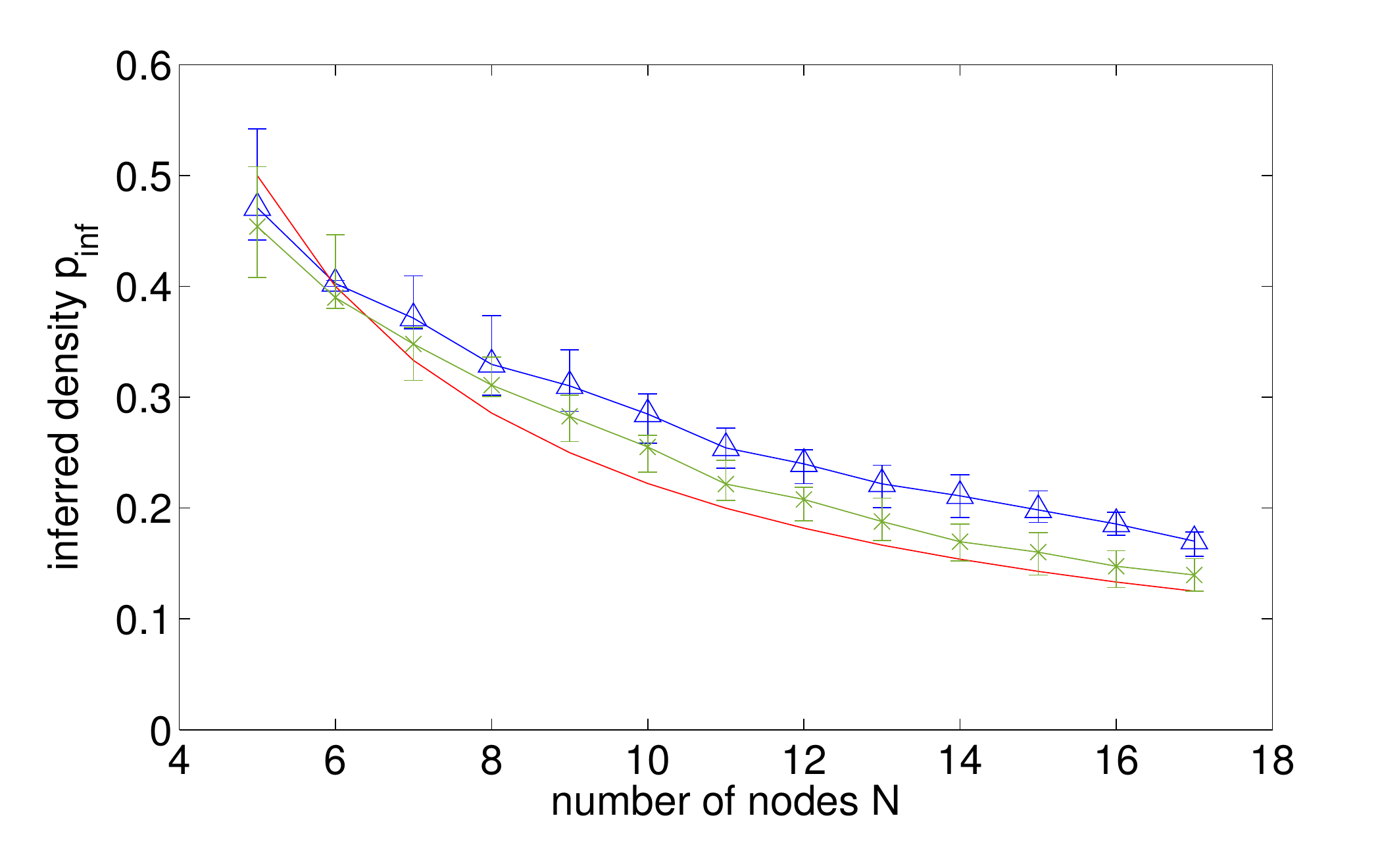}
   \end{center}
   \caption{\label{Fig:InfDens}(Color online) Average inferred link
     densities $p_{inf}$ as a function of network size, if our maximum
     entropy method is applied using unthresholded estimated mutual
     informations. The actual network densities are $p=0.1$ (blue
     triangles), and $p=1$ (green crosses). The red curve is the
     density of a network in which every node interacts with two
     neighbors on average, and given for comparison. The inferred
     density is calculated as the number of inferred links over the
     number of possible links, ${N \choose 2}$. Each curve is
     generated from 100 realizations of natural frequencies and
     adjacency matrices with given density $p$. The error bars are
     given by the $25\%$ and $75\%$ quantiles.}
\end{figure}

The presented maximum entropy estimator calculates CMI between each
pair of oscillators conditioned on every other oscillator from
supplied constraints on the estimated mutual information and
univariate entropies. A link is inferred if the CMI between the two
oscillators is nonzero. However, we find, for a given system size, the
average inferred density doesn't depend much on the actual density of
the network. Figure~\ref{Fig:InfDens} shows the maximum density for
varying network sizes. If a network of higher density is analyzed, the
method can still be expected to infer existing links, however it will
fail to identify a significant number of true links. This is a result
of the method having at least one zero conditional information for all
subsets of variables greater than two, so for example this method will
not infer any triangles, see Sec.~\ref{SSec:ProofN3}. If in contrast a
network is analyzed that is sparser than the average density inferred
by the method, our findings necessitates the use of a threshold to
reduce the detection of spurious links.

To speed up the optimization we use a strictly stronger set of
inequalities here, where it is assumed that every atom is
non-negative. This provides a lower bound for the maximum entropy.  If
interactions are truly described by bivariate interactions only, then
negative atoms are expected to be negligible, as they would indicate
higher-order interactions. This approximation should only be employed
if pairwise interactions have already been established as a good
model, which we tested for the Kuramoto model
in~\cite{Martin2015::PRL}.  Numerical comparisons at smaller system
sizes have indicated that this is indeed a viable approximation. To
that end 100 realization with a length of $T=10,000$ were evaluated at
a system size of $N=9$ using both the exact constraints and the
approximate ones. The biggest relative error of the approximate
maximum entropy estimate was $0.087\%$.

%%%%%%% FIG %%%%%%%%%%%%%%%%%%%%%%%%%%%%%%%%%%%%%%%%%%%%%%%%%%%%%%%%%
%%%%% FIG %%%%%%%%%%%%%%%%%%%%%%%%%%%%%%%%%%%%%%%%%%%%%%%%%%%%%%%%%

\begin{figure}[!h]
%%\begin{figure}[htbp]
  \begin{center}
    \includegraphics*[width=\columnwidth]{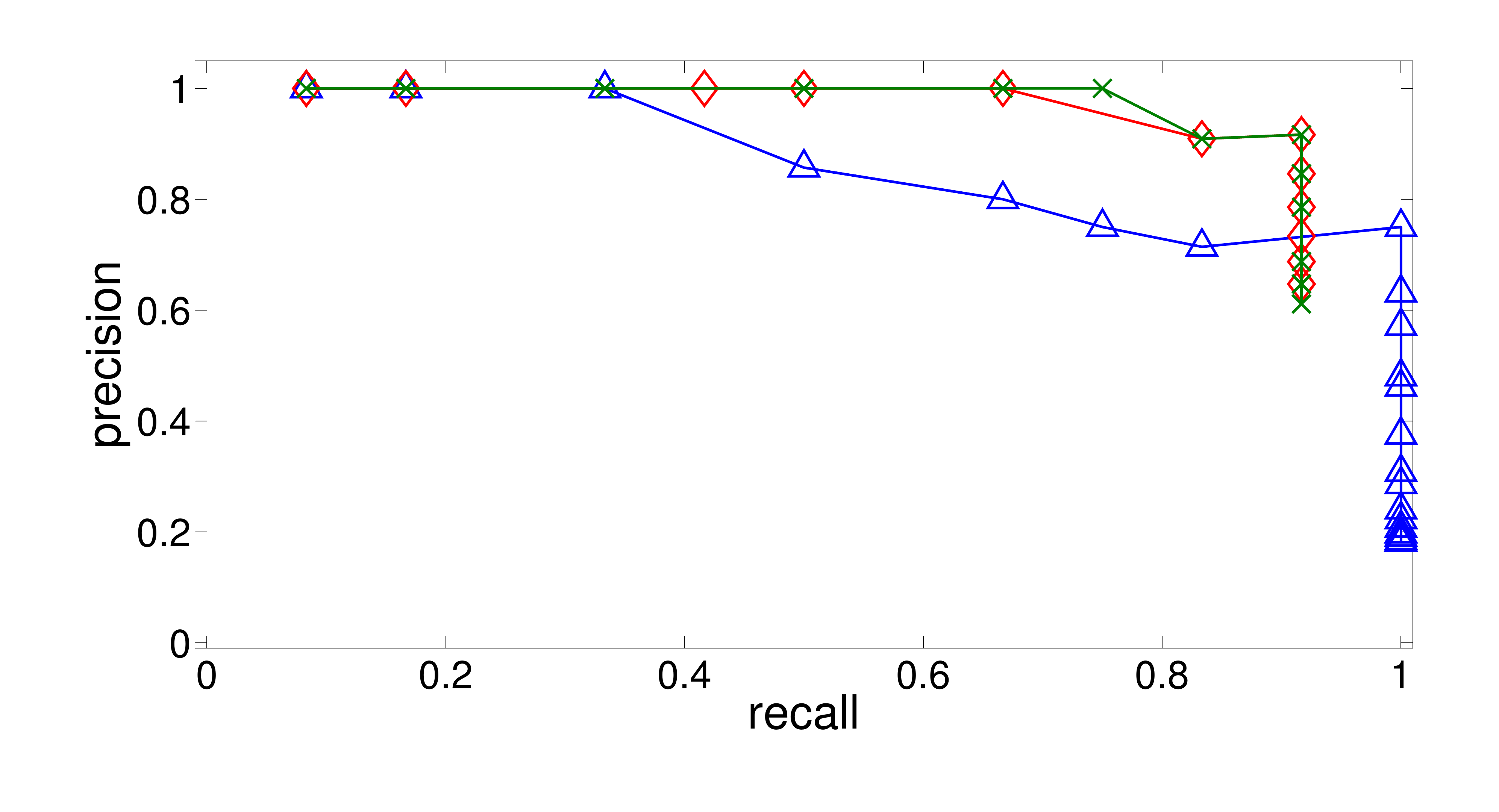}
   \end{center}
   \caption{\label{Fig:PrecRec}(Color online) Representative
     precision/recall curves: Thresholding of MI matrix alone, and not
     using any maximum entropy method (blue triangles); thresholding
     of CMI matrix obtained using our maximum entropy method on the
     (unthresholded) mutual informations (green crosses); thresholding
     of mutual informations, and then using the maximum entropy method
     (red diamonds).}
\end{figure}

For global thresholding there are two obvious options: either
threshold the mutual informations and then apply our maximization
procedure, or apply our maximum entropy method first and then threshold the CMI
matrix. Our assessment of the method's performance is based on the
precision (ratio of correctly inferred links to all the inferred
links) and the recall (ratio of correctly inferred links over existing
links) (see for example~\cite{rijsbergen1979information}).  As shown
in Fig.~\ref{Fig:PrecRec}, using these valuation metrics neither
approach seems to be superior over the other. Both ways generally
improve the performance of merely thresholding the mutual information
without using any maximum entropy method at all.

%%%%%%% FIG %%%%%%%%%%%%%%%%%%%%%%%%%%%%%%%%%%%%%%%%%%%%%%%%%%%%%%%%%
%%%%% FIG %%%%%%%%%%%%%%%%%%%%%%%%%%%%%%%%%%%%%%%%%%%%%%%%%%%%%%%%%

\begin{figure}[!h]
%%\begin{figure}[htbp]
  \begin{center}
    \includegraphics*[width=\columnwidth]{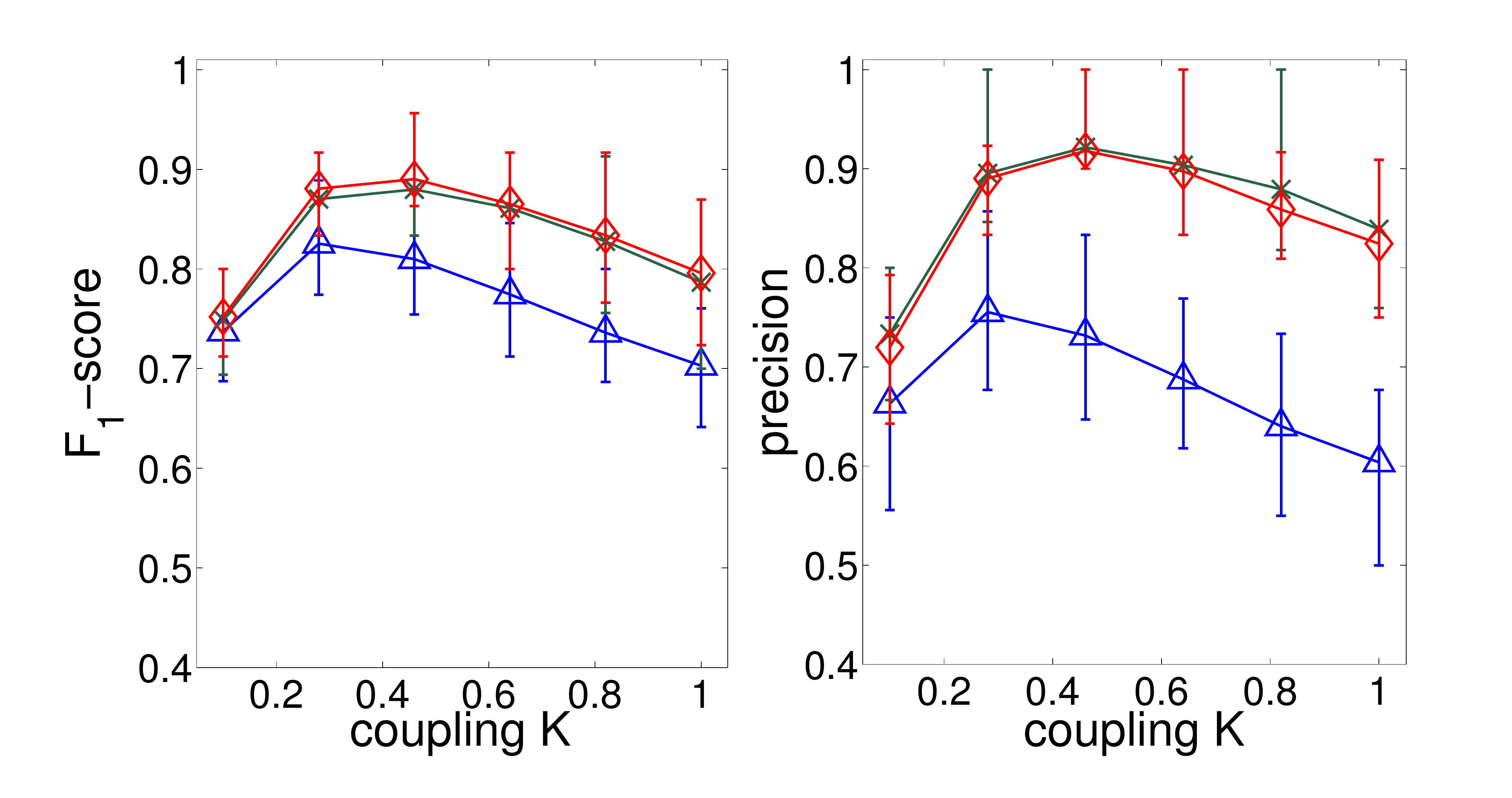}
   \end{center}
   \caption{\label{Fig:BestFscore}(Color online) For different
     thresholding methods the global threshold has been picked that
     leads to the highest $F_1$-score (left). The precision
     corresponding to that threshold is also plotted for comparison
     (right). The symbols are the same as in
     Fig.~\ref{Fig:PrecRec}. The process has been applied to an
     ensemble of 100 realizations and the averages are shown. Error
     bars are given by the $25\%$ and $75\%$ quantiles.}
\end{figure}

To make this observation more quantitative, it is useful to have a
single real number valuation metric to compare performances. We have
chosen the $F_1$-score~\cite{rijsbergen1979information}, defined as
$F_1=2\cdot \frac{precision\cdot recall}{precision + recall}$, because
it treats precision and recall symmetrically and it is not clear that
either measure should be preferred in a general context. From
Fig.~\ref{Fig:BestFscore} it is apparent that the best performance is
achieved at $K \approx 0.5$.  Considering the coupling is given as
$\frac{K}{N}=\frac{0.5}{12}\approx 0.042$ which is of the same order
of magnitude as the noise $G=0.05$, this indicates that our method
performs particularly well in the weak coupling regime where no
oscillators are synchronized.

%%%%%%% FIG %%%%%%%%%%%%%%%%%%%%%%%%%%%%%%%%%%%%%%%%%%%%%%%%%%%%%%%%%
%%%%% FIG %%%%%%%%%%%%%%%%%%%%%%%%%%%%%%%%%%%%%%%%%%%%%%%%%%%%%%%%%

Figure~\ref{Fig:BestFscore} also shows that a generally higher precision
and $F_1$-score can be achieved using our method as an additional step
after the mutual information thresholding, only partially compromising
the recall. The problem of finding a suitable global threshold that
actually achieves that performance remains open. In the following
section we outline a surrogate based method of finding a non-global
threshold that displays a performance comparable to the global
thresholding discussed above.

\subsubsection{Finding significance thresholds}

A problem for the method's performance on the Kuramoto model is posed
by the fact that two disconnected nodes can have a high estimated
mutual information in a finite time series, if their effective
frequencies are close to each other.  To account for this we generate
surrogates that preserve these effective frequencies as well as the
oscillator's autocorrelations.  First the effective frequencies are
removed from the time series, subtracting from each oscillator the
linear function that interpolates between the initial and final value
of their unwrapped phase. That way each oscillator's time series
begins and ends at the same value.  In the next step, the Iterative
Amplitude Adapted Fourier Transform Algorithm
(IAAFT)~\cite{Schreiber1996::PRL} is applied to the trend-removed
time series. As a last step, the trends are added back in and for each
oscillator a random number uniformly drawn from $0$ to $2\pi$ is added
to every value of the time series. This corresponds to randomizing the
initial conditions.  The mutual informations between the so obtained
time series are estimated in the same way as before. This provides an
estimate for the mutual information for each pair of oscillators that is
not due to their coupling.

To obtain a (local) threshold, a statistical significance level has to be
chosen.  Since higher significance levels require more surrogate
series, the problem can become computationally very
expensive. In~\cite{Margolin2006::BMCbioinfo} they suggest that good
performance can be expected in the regime of $q\approx 98.5\%$,
because there are $12 \choose {2}$ $ = 66$ possible links in our
system and $\frac{1}{66} \approx 1.5\%$. The rational behind this is
pick $q$ so that we expect to keep on average one false link with this
threshold.

%%%%%%% FIG %%%%%%%%%%%%%%%%%%%%%%%%%%%%%%%%%%%%%%%%%%%%%%%%%%%%%%%%%
\begin{figure}[!h]
%%\begin{figure}[htbp]
  \begin{center}
    \includegraphics*[width=\columnwidth]{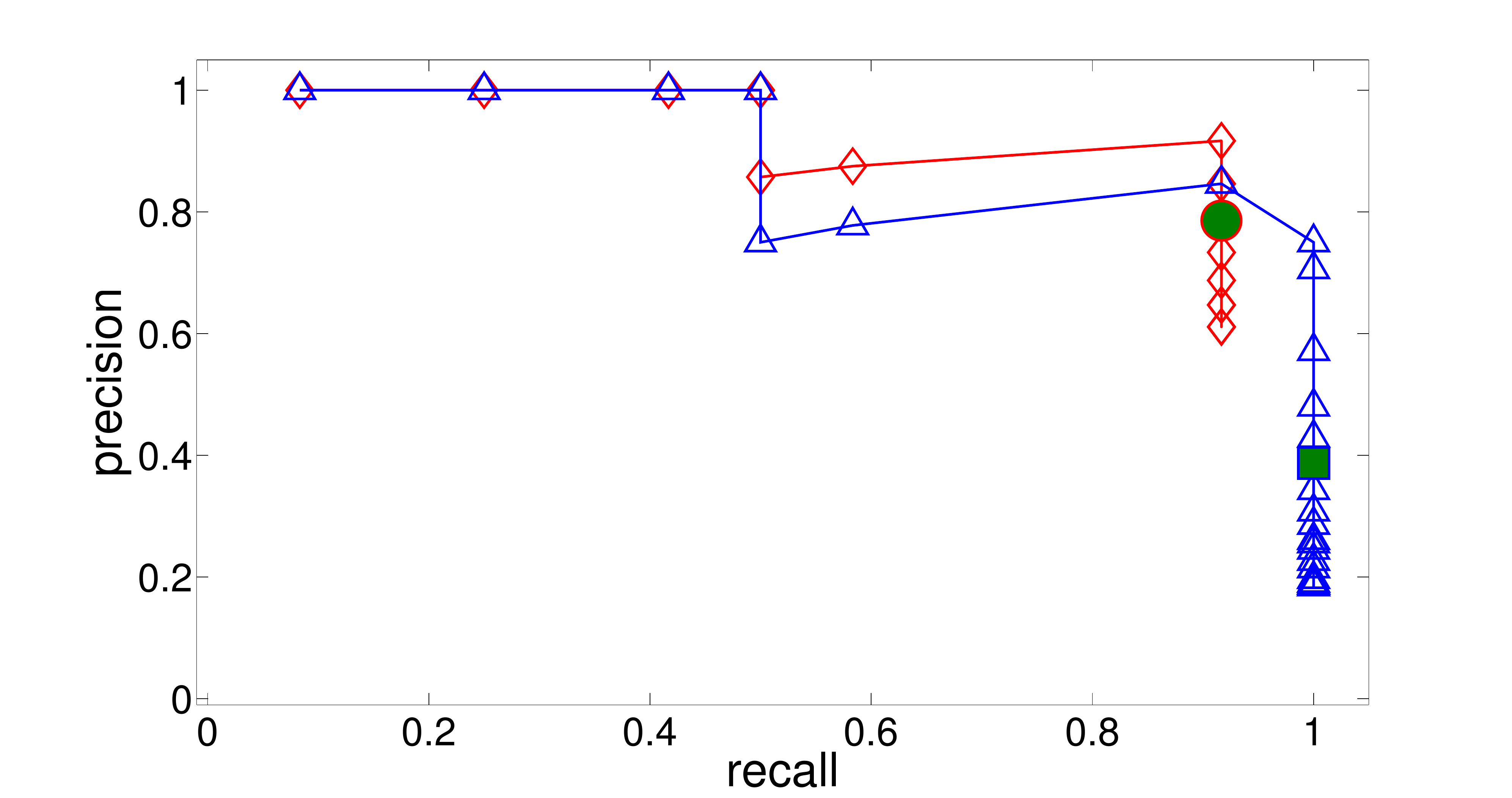}
  \end{center}
  \caption{\label{Fig:NonGlobalThres}(Color online) A single typical
    realization of a precision/recall curve for only thresholding the
    mutual information (blue triangles) and the maximum entropy method
    being applied to thresholded mutual informations (red diamonds),
    similar to Fig.~\ref{Fig:PrecRec}. The blue square filled green
    shows the performance of applying local thresholds to the MI
    matrix and the red circle filled green the performance of applying
    the method to the locally thresholded mutual informations. The
    thresholds are determined by the surrogate method discussed in the
    main text using the $q=98.5$ percentile. 700 surrogates were
    generated.}
\end{figure}
%%%%% FIG %%%%%%%%%%%%%%%%%%%%%%%%%%%%%%%%%%%%%%%%%%%%%%%%%%%%%%%%%

As Fig.~\ref{Fig:NonGlobalThres} indicates, the surrogate-based local
thresholding method achieves good performance after our maximum
entropy method is applied. This claim is substantiated by statistics
as shown in Table~\ref{Tab:NonGlobalEns}, clearly establishing the
benefit of our maximum entropy method.
% Even though there exists a global
% threshold for which the relevance networks would outperform the
% presented approach, it is not clear how to pick such a threshold in
% general.

\begin{table}[]
\centering
\begin{tabular}{|l|l|l|l|}
  \hline
  & precision  & recall  & $F_1$-score \\ \hline
  mutual informations only  & 0.369 & 0.997  & 0.534 \\ \hline
  with maximum entropy method & 0.772  & 0.834  & 0.789 \\ \hline
\end{tabular}
\caption{\label{Tab:NonGlobalEns}
  Performance of local thresholding averaged over 100 realizations
  with $T=10,000$ and $q=90\%$ using 100 surrogates each. The CMI
  achieved a higher $F_1$-score in all but 3 cases. The average
  difference in performance was $\Delta F_1 = 0.255 (+0.079, -0.045)$
  with $25\%$ and $75\%$ quantiles given.}
\end{table}

\section{Discussion \& Conclusions}

% {\bf D1: Showed we could extend the maximum entropy scheme introduced
%   in~\cite{***PRL} so a wide range of information theoretic quantities
%   can be conditioned on. We also extended our maximum entropy scheme
%   to be able to use continuous variables. Allows our technique to be
%   applied to a much larger range of cases. There are pathological
%   problems for linear optimization that will scale exponentially with
%   $N$, but there will always be a slightly perturbed problem that will
%   scale polynomially~\cite{Vershynin2009::SJC}}

In this paper we extended the method we introduced
in~\cite{Martin2015::PRL} to compute the maximum entropy conditioned
on a wide range of information-theoretic quantities --- beyond the
bivariate mutual informations and univariate entropies --- using
linear optimization. We have also shown how to implement our method
with continuous variables, no longer limiting it to discrete ones,
making our technique applicable to a much broader range of
problems. While there are pathological linear optimization problems
whose running time will scale exponentially with the number of
variables, $N$, there will always be a slightly perturbed problem such
that our method will scale polynomially~\cite{Vershynin2009::SJC}.

% {\bf D3: Both methods are nonparametric, in that they do not generate
%   a probability distribution. While this may result in an
%   unsatisfaiable diagram (violating non-Shannon inequalities), we
%   showed that in the common case where the maximum diagram has only
%   non-negative atoms there will be a set of variables that satisfy it
%   --- in fact there will be an infinite number of such sets....***}

Our method is nonparametric in that it does not generate a
corresponding probability distribution. This may result in a diagram
for which no probability distribution can be constructed (since it may
violate a non-Shannon inequality). However, we proved in
Sec.~\ref{SSec:ConstructPositivity} that in the common case where the
maximum diagram has only non-negative regions it will indeed be
satisfiable.

% {\bf D4: Showed that our methods perform well in the undersampled
%   regime, where other techniques fail. This means our method can be
%   applied to systems with many states, that haven't been fully
%   sampled. Showed that our methods offer computational speedups over
%   competeing techniques --- especially when the variables have
%   cardinality greater than 2. Method independent of cardinality of
%   variables. This makes our techniques perfectly positioned to analyze
%   large systems where computation time can be prohibitive using
%   current techniques, and the size of the phase space prohibits
%   accurate samplinge.}

Since our methods do not require the direct estimate of any
probability distribution, we can apply them in the undersampled
regime. We demonstrated in Sec.~\ref{SSec:UnderSamp} that in this
regime our method offers a much more accurate estimate of the maximum
entropy.  Additionally, in Sec.~\ref{SSec:ComputTime} we demonstrated
that our method offers computational speedups over competing
techniques when the variables have cardinality greater than 2.  This
makes our techniques perfectly positioned to analyze systems of larger
cardinality variables where the size of the phase space can make both
computation time and accurate sampling prohibitive.

% {\bf D5: Our technique motivated us to introduce a non-parametric
%   formulation of connected informations. This formulation can be
%   computed using our technique, and so can be applied to larger
%   systems, and in the undersampled regime. We showed that in cases of
%   interest it will give the same results as the parametric
%   formulation}

Motivated by our new ability to easily compute the maximum entropy
given information-theoretic constraints, we introduced a nonparametric
formulation of connected informations,
Sec.~\ref{SSec:ConectInfo}. This can be computed directly using our
linearly optimized maximum entropy, and hence has its computational
and sampling advantages. For paradigmatic examples of higher-order
relationships --- which connected informations attempt to detect ---
we demonstrated that our nonparametric method will give the same
result as the standard one, Sec.~\ref{SSSec:Examp}.

We have also expanded on our work in~\cite{Martin2015::PRL}, where we
have now analyzed two resting-state human brain networks. It is highly
desirable to know if these networks can be accurately described with
pairwise measurements, as it would tremendously simplify their
analysis, and is common practice. Previous results indicated that this
is the case, but only when the signal is
binarized~\cite{Watanabe2013::NatCom}. In both networks analyzed we
have shown that conditioning on the first two moments of the
distributions exhibits a marked sensitivity to the number of states
the system is discretized to, Sec.~\ref{SSec:RestNet}. On the other
hand our method appears to be robust to the specific discretization,
as was also seen in~\cite{Martin2015::PRL} for the case of the
Kuramoto model. This indicates that pairwise measurements can still
capture the vast majority of the complexity of these networks, but
only when nonlinear relationships are taken into account.

Finally, we used our entropy maximization method to infer conditional
mutual informations, and hence to infer the structural connectivity of
a network, Sec.~\ref{SSec:NetInf}.  We showed that our maximum entropy
estimation can be used to improve on the performance of naively
thresholding the mutual informations to infer networks of a dynamical
system. For the Kuramoto model it was also evident that our method
performs particularly well in the weak coupling regime, where other
methods do struggle. For example, the main method proposed in
Ref.~\cite{Tirabassi2015::SR} achieved its best results for the
Kuramoto model in the strong coupling regime.

We also managed to demonstrate that our particular thresholding method
achieves high precision, while also retaining higher recall than, for
example, in Ref.~\cite{Margolin2006::BMCbioinfo}.  There they used a
method similar in spirit to ours, where they estimated the network
based on the thresholded mutual information between all pairs of
variables, as well as set the weakest mutual information between every
triplet of variables to zero. While they justified this with the data
processing inequality, we showed in Sec.~\ref{SSec:ProofN3} that this
also can justified as a result of maximum entropy estimation, giving
further credence to their method. It must be noted however, that
bigger system sizes and higher link densities were considered in
Ref.~\cite{Margolin2006::BMCbioinfo} than can be treated with our
presented method.

In conclusion, we have shown that our entropy maximization performs
well in the undersampled regime, and for high cardinality
variables. This helps resolve two outstanding problems with maximum
entropy estimation, as noted in~\cite{Yeh2010::Ent}. We have also
shown that this method can be applied to real world problems facing
researchers, using fMRI data and network inference as examples. While
we have given a few obvious applications for our method, given its
broad nature it is our belief that many researchers will find uses for
it that we have yet to anticipate.

This project was financially supported by NSERC (EM and JD) and by the
Czech Science Foundation project GA13-23940S and the Czech Health
Research Council project NV15-29835A (JH). AM was financially
supported by the DAAD. All authors would like to thank the MPIPKS for
its hospitality and hosting the international seminar program
``Causality, Information Transfer and Dynamical Networks'', which
stimulated some of the involved research. We also would like to thank
P. Grassberger for many helpful discussions.

% \bibliographystyle{unsrt} 
% \bibliography{MaxEntPRE.bib}

\begin{thebibliography}{10}

\bibitem{Jaynes1957::PRa}
Edwin~T Jaynes.
\newblock Information theory and statistical mechanics.
\newblock {\em Phys. Rev.}, 106(4):620, 1957.

\bibitem{Schneidman2006::Nat}
Elad Schneidman, Michael~J Berry, Ronen Segev, and William Bialek.
\newblock Weak pairwise correlations imply strongly correlated network states
  in a neural population.
\newblock {\em Nature}, 440(7087):1007--1012, 2006.

\bibitem{Lezon2006::PNAS}
Timothy~R Lezon, Jayanth~R Banavar, Marek Cieplak, Amos Maritan, and Nina~V
  Fedoroff.
\newblock Using the principle of entropy maximization to infer genetic
  interaction networks from gene expression patterns.
\newblock {\em Proc. Natl. Acad. Sci.}, 103(50):19033--19038, 2006.

\bibitem{Wood2012::PNAS}
K.~Wood, S.~Nishida, E.~D. Sontag, and P.~Cluzel.
\newblock Mechanism-independent method for predicting response to multidrug
  combinations in bacteria.
\newblock {\em Proc. Natl. Acad. Sci.}, 109(30):12254--12259, 2012.

\bibitem{Schneidman2003::PRL}
Elad Schneidman, Susanne Still, Michael~J Berry, William Bialek, et~al.
\newblock Network information and connected correlations.
\newblock {\em Phys. Rev. Lett.}, 91(23):238701, 2003.

\bibitem{Stephens2010::PRE}
G.~J. Stephens and W.~Bialek.
\newblock Statistical mechanics of letters in words.
\newblock {\em Phys. Rev. E}, 81:066119, Jun 2010.

\bibitem{Cover2006}
T.~M. Cover and J.~A. Thomas.
\newblock {\em Elements of information theory}.
\newblock John Wiley \& Sons, 2006.

\bibitem{Nemenman2011::Ent}
Ilya Nemenman.
\newblock Coincidences and estimation of entropies of random variables with
  large cardinalities.
\newblock {\em Entropy}, 13(12):2013--2023, 2011.

\bibitem{Martin2015::PRL}
E.~A. Martin, J.~Hlinka, and J.~Davidsen.
\newblock Pairwise network information and nonlinear correlations.
\newblock Submitted to Phys. Rev. Lett.

\bibitem{Yeh2010::Ent}
F.~C. Yeh, A.~Tang, J.~P. Hobbs, P.~Hottowy, W.~Dabrowski, A.~Sher, A.~Litke,
  and J.~M. Beggs.
\newblock Maximum entropy approaches to living neural networks.
\newblock {\em Entropy}, 12(1):89--106, 2010.

\bibitem{Watanabe2013::NatCom}
T.~Watanabe, S.~Hirose, H.~Wada, Y.~Imai, T.~Machida, I.~Shirouzu, S.~Konishi,
  Y.~Miyashita, and N.~Masuda.
\newblock A pairwise maximum entropy model accurately describes resting-state
  human brain networks.
\newblock {\em Nat. Commun.}, 4:1370, 2013.

\bibitem{Timme2014}
M.~Timme and J.~Casadiego.
\newblock Revealing networks from dynamics: an introduction.
\newblock {\em J. Phys. A Math. Theor.}, 47(34):343001, 2014.

\bibitem{Eguiluz2005::PRL}
V.~M. Eguiluz, D.~R. Chialvo, G.~A Cecchi, M.~Baliki, and A.~V. Apkarian.
\newblock Scale-free brain functional networks.
\newblock {\em Phys. Rev. Lett.}, 94(1):018102, 2005.

\bibitem{Margolin2006::BMCbioinfo}
A.~A. Margolin, I.~Nemenman, K.~Basso, C.~Wiggins, G.~Stolovitzky, R.~D Favera,
  and A.~Califano.
\newblock Aracne: an algorithm for the reconstruction of gene regulatory
  networks in a mammalian cellular context.
\newblock {\em BMC Bioinformatics}, 7(Suppl 1):S7, 2006.

\bibitem{Runge2015::NC}
J.~Runge, V.~Petoukhov, J.~F. Donges, J.~Hlinka, N.~Jajcay, M.~Vejmelka,
  D.~Hartman, N.~Marwan, M.~Palu{\v{s}}, and J.~Kurths.
\newblock Identifying causal gateways and mediators in complex spatio-temporal
  systems.
\newblock {\em Nat. Commun.}, 6, 2015.

\bibitem{Tirabassi2015::SR}
G.~Tirabassi, R.~Sevilla-Escoboza, J.~M Buld{\'u}, and C.~Masoller.
\newblock Inferring the connectivity of coupled oscillators from time-series
  statistical similarity analysis.
\newblock {\em Sci. Rep.}, 5, 2015.

\bibitem{Frenzel2007::PRL}
S.~Frenzel and B.~Pompe.
\newblock Partial mutual information for coupling analysis of multivariate time
  series.
\newblock {\em Phys. Rev. Lett.}, 99(20):204101, 2007.

\bibitem{Runge2015}
J.~Runge and J.~Davidsen.
\newblock Continuous random variables and time graphs.
\newblock In Preperation.

\bibitem{Yeung2008information}
Raymond~W Yeung.
\newblock {\em Information theory and network coding}.
\newblock Springer, 2008.

\bibitem{Darroch1972::AnMatStat}
J.~N. Darroch and D.~Ratcliff.
\newblock Generalized iterative scaling for log-linear models.
\newblock {\em Ann. Math. Stat.}, 43(5):1470--1480, 1972.

\bibitem{Shirer2011}
W.~R. Shirer, S.~Ryali, E.~Rykhlevskaia, V.~Menon, and M.~D. Greicius.
\newblock Decoding subject-driven cognitive states with whole-brain
  connectivity patterns.
\newblock {\em Cereb. Cortex}, 2011.

\bibitem{Hlinka2011Neuroimage}
J.~Hlinka, Milan Palu{\v{s}}, M.~Vejmelka, D.~Mantini, and M.~Corbetta.
\newblock {Functional connectivity in resting-state fMRI: Is linear correlation
  sufficient?}
\newblock {\em NeuroImage}, 54:2218--2225, 2011.

\bibitem{butte2000mutual}
A.~J. Butte and I.~S. Kohane.
\newblock Mutual information relevance networks: functional genomic clustering
  using pairwise entropy measurements.
\newblock In {\em Pac. Symp. Biocomput.}, volume~5, pages 418--429. World
  Scientific, 2000.

\bibitem{kuramoto1975self}
Y.~Kuramoto.
\newblock Self-entrainment of a population of coupled non-linear oscillators.
\newblock In {\em International symposium on mathematical problems in
  theoretical physics}, pages 420--422. Springer, 1975.

\bibitem{Erdos1960}
P.~Erd{\"o}s and A.~R{\'e}nyi.
\newblock On the evolution of random graphs.
\newblock {\em Publ. Math. Inst. Hung. Acad. Sci}, 5:17--61, 1960.

\bibitem{rijsbergen1979information}
C.~J. Rijsbergen.
\newblock {\em Information retrieval. online book
  http://www.dcs.gla.ac.uk/Keith/Chapter.7/Ch.7.html}.
\newblock Butterworth-Heinemann, 1979.

\bibitem{Schreiber1996::PRL}
T.~Schreiber and A.~Schmitz.
\newblock Improved surrogate data for nonlinearity tests.
\newblock {\em Phys. Rev. Lett.}, 77(4):635, 1996.

\bibitem{Vershynin2009::SJC}
R.~Vershynin.
\newblock Beyond hirsch conjecture: Walks on random polytopes and smoothed
  complexity of the simplex method.
\newblock {\em SIAM J. Comput.}, 39(2):646--678, 2009.

\end{thebibliography}

\end{document}